\documentstyle[twocolumn,prl,aps]{revtex}

\input  epsf.sty   

\begin {document}

\draft

\title
{
Identification of Non-unitary triplet pairing 
in a heavy Fermion superconductor UPt$_3$
} 

\author{Kazushige Machida$^{1,\ast}$, and Tetsuo Ohmi$^{2,\dagger}$}

\address{$^1$Department of Physics, Okayama University,
         Okayama 700, Japan\\ 
     $^2$Department of Physics, Kyoto University, Kyoto 
606-01, Japan\\\vspace{.5cm}
\parbox{14cm}{\rm
\hspace{.3cm}
 A NMR experiment recently done by Tou {\it et al.} on a heavy Fermion 
superconductor UPt$_3$ is interpreted in terms of 
a non-unitary spin-triplet pairing state which we have been 
advocating. The proposed state successfully explains various 
aspects of the seemingly complicated Knight shift 
behaviors probed for major orientations, including a remarkable 
{\bf d}-vector rotation under weak fields. This entitles UPt$_3$
as the first example that a charged many body system forms a spin-triplet odd-parity pairing at low temperatures and demonstrates unambiguously that the putative 
spin-orbit coupling in UPt$_3$ is weak.
}}

\maketitle

Except for the famous example of neutral superfluid Fermion 
system $^3$He, no charged Fermion many body system has been known to
exhibit spin triplet pairing in nature so far; that is, a triplet superconductor is not firmly identified yet.
A heavy Fermion superconductor UPt$_3$\cite{heffner} has been 
regarded as a prime candidate for a triplet superconductor because the observed double transition ($T_{c1}\sim$0.58K and 
$T_{c2}\sim$0.53K)
and multiple phases (A, B, and C phases) in $H$ vs $T$ apparently point to some unconventional pairing state where a Cooper pair must have internal degrees of freedom responsible 
for the multiple phases.

A central controversy in identifying the pairing symmetry in UPt$_3$
lies in whether the internal degrees of freedom as the origin nearly degenerate superconducting transition temperatures  come 
either from the orbital part\cite{sauls,joynt} or from the spin part\cite{our} of a Cooper pair.
There are the three major scenarios for explaining the multiple 
phase diagram, apart from the so-called accidental degeneracy scenario\cite{garg}
whose proposed state is not qualified in detail as yet, being 
unable to
compare it with the above three on an equal footing. 
The orbital scenarios identify the pairing symmetry as a two-dimensional representation whose multi-dimensionality of the 
orbital part is responsible for the degeneracy, namely the singlet 
pairing E$_{1g}$ proposed by Joynt\cite{joynt}, the triplet pairing 
E$_{2u}$ by Sauls\cite{sauls}. The latter assumes that the degeneracy 
is broken by the antiferromagnetism and spin-orbit coupling (SOC) is strong. Thus in E$_{2u}$ the spin of a pair is firmly locked to the 
crystal lattice. 
In the spin scenario\cite{our}, 
on the other hand, the degeneracy comes from the spin part of the pair in which weak SOC is assumed by considering that SOC 
in a one-particle level is already taken account in forming 
heavily renormalized quasi-particle characteristic of heavy Fermion materials.

It was not an easy task to distinguish the above three 
scenarios.
Since usual thermodynamic or transport measurements extract
power low indices in their $T$-dependence, which reflect the energy gap topology. The accumulated experiments all point to the 
gap vanishing at both point and line\cite{heffner}. This is consistent with the above scenarios which yield more or less similar gap topology. 
An obvious and decisive experiment is to measure the Pauli spin susceptibilities $\chi_i$ for the principal orientations ($i=a,b, c$-axes)
in the hexagonal crystal since in the singlet E$_{1g}$, $\chi_i$ decreases
for all directions, and in the triplet E$_{2u}$, $\chi_a$ and $\chi_b$ ($\chi_c$) remain  unchanged (decreases) below $T_{c}$, irrespective of the 
three phases (A, B, and C), that is, the three phases give rise 
to the identical behavior in their scenarios. In the spin scenario the $\chi_i$ behavior is much more subtle and complicated  as shown shortly.

Owing to technological developments in making high quality single
crystals by ${\bar{\rm O}}$nuki group and improvements of NMR instrumentation by Asayama and Kitaoka group, their joint work\cite{nmr} has culminated
in success of measuring the Knight shift (KS) by probing $^{195}$Pt nuclear magnetic resonance using single crystals (two crystals: $\sharp$3, $\sharp$4).
They extend and refine the previous experiment\cite{tou} to much 
more wider fields and yet unexplored field orientations. The new experiment covers virtually all the phases in $H$ vs $T$ plane for major $H$ orientations, enabling us to completely determine 
the Cooper pair spin structure.

The purpose of this paper is to demonstrate that a single and simple Ginzburg-Landau (GL) free enegy based on a few basic 
assumptions can explain all aspects of the seemingly complicated results of their experiment\cite{nmr}
summarized in Table 1, leading us to 
identify the pairing symmetry in UPt$_3$.
Before going into detailed analysis of the Knight shift changes
for major orientations, we first reexamine
our fundamental assumptions
in the light of the new NMR experiment.

The first and most central assumption, that is, 
the effective SOC felt by Cooper pairs  is relatively 
weak. This implys that under symmetry operations in D$_{6h}\times SO(3)\times{\cal T}\times U(1)$ (D$_{6h}$ hexagonal point group, $SO(3)$ spin 
rotation, ${\cal T}$ time reversal and $U(1)$ gauge symmetry) 
the spin and orbital parts of the spin-triplet order parameter described by $\Delta_{\alpha,\beta}({\bf k})=i({\bf \sigma}\cdot{\bf 
d}({\bf k})\sigma_2)_{\alpha,\beta}$ in terms of the {\bf d} vector behave independently.
The strong SOC assumption adopted by the orbital scenarios, under the symmetry operations the spin and orbital part transform simultaneously and the Cooper pair spin is strongly locked to the crystal lattice and is not rotatable. In this strong 
SOC case  one must resort to the remaining possibility of the orbital part, which is the case in the E$_{2u}$ 
scenario. Tou's result\cite{nmr} that the field as low as $H (\parallel {\bf c})\sim$2kG makes the Cooper pair spin rotate clearly shows 
that weak SOC is indeed the case for UPt$_3$. (Note that the BC boundary is far off at $\sim$12kG.) This is one of the most important significance in their experiment because  
almost all theories for  heavy  
Fermion superconductors take strong SOC for granted since Anderson's proposal\cite{anderson}.

The second assumption of spin-triplet degeneracy of our pairing 
function is consistent with Tou' experiment\cite{nmr}
since through their 
Knight shift change one can estimate the Pauli spin susceptibilities $\chi_i$ for various directions coming from the quasi-particles 
at the Fermi surface. (Don't confuse those with the total bulk susceptibilities which are of course very anisotropic as is well known)  According to their data\cite{nmr}, $\chi_i$ both parallel and 
perpendicular to the $c$-axis are  nearly isotropic, supporting the triple degeneracy in spin space and leading to  nearly $SO(3)$ 
symmetry.

The third assumption that the antiferromagnetic (AF) fluctuations 
characterized by the triple-$Q$ vectors with ${\bf Q}_1=({1\over 
2}, 0,0)$ and its equivalent positions ${\bf Q}_2$ and ${\bf Q}_3$ in reciprocal space 
($a^{\ast}=b^{\ast}={4\pi\over a\sqrt3}$, $c^{\ast}={2\pi\over c}$) in 
the hexagonal plane\cite{lussier} are responsible for breaking the triple degeneracy. This is not inconsistent with their experiment because the static AF order is not present~\cite{tou}. The triple 
$Q$-AF fluctuation structure probed by neutron experiments\cite{aeppli3,isaacs,lussier} is symmetry-broken from 
the outset by, for example,
the incommensurate lattice modulation whose 
presence or absence exhibit a good correlation with the double 
transition\cite{midgley}, leading to a different weight to the 
triple $Q$ fluctuations. Therefore it is possible to break the hexagonal symmetry by this longitudinal AF fluctuations. Without loosing generality we take ${\bf Q}_1\parallel {\bf a}^{\ast}\parallel {\bf b}$ whose fluctuation is differed from those at 
${\bf Q}_2$ and ${\bf Q}_3$, implying that the symmetry of the hexagonal plane is lowered to two-fold symmetry, which is 
indeed observed by Tou's experiment\cite{nmr}
as will be seen. They single 
out the two-fold unique axis in the nominally hexagonal plane. That is remarkable.

Let us now start out by writing down the phenomenological 
GL free energy, derived based on the above three assumptions:

$$F=\alpha_0(T-T_{c0})|{\bf d}|^2+{1\over 2}\beta_1|{\bf d}|^4
+{1\over 2}\beta_2|{\bf d}^2|^2-\gamma|{\bf b}\cdot{\bf d}|^2-\lambda|{\bf c}\cdot{\bf d}|^2$$

\noindent
where ${\bf Q}_1\parallel {\bf b}$. The last term $\lambda|{\bf c}\cdot{\bf d}|^2$ which is somewhat ad hoc
expresses a weak anisotropy of the 
order parameter in spin space, reflecting the fact that the Knight shift 
changes below T$_c$ for $H$=2kG parallel to the $c$-axis\cite{nmr}.
This free energy leads to

$$F=\sum_{j=a,b,c}\alpha_0(T-T_{c}^j)|d_j|^2+{1\over 2}\beta_1|{\bf d}|^4+{1\over 2}\beta_2|{\bf d}^2|^2$$

\noindent
with $T^b_c=T_{c0}+{\gamma\over\alpha_0}>
T^c_c=T_{c0}+{\lambda\over\alpha_0}>T^a_c$ for 
$\gamma>\lambda>0$. 
The three transition temperatures $T^a_c$, $T^b_c$ and $T^c_c$ which were originally degenerate are split into the two groups: 
$T^b_c$ and \{$T^a_c$ and $T^c_c$\} by the symmetry breaking field as mentioned. The latter two are further 
assumed to be slightly 
different $T^a_c<T^c_c$ due to the $\lambda$-term because of the small  uniaxial anisotropy of the system.

At $T=T^b_c$, which is identified as the upper critical temperature  $T_{c1}=0.58K$ the A phase characterized by $d_a=0, 
d_b\neq 0$ and $d_c=0$ appears first.
Then at a lower temperature, $T=T^c_c$ which is identified as $T_{c2}=0.53K$ 
the second order transition from the A phase to the B phase characterized by a non-unitary triplet state: 
$d_a=0, d_b\neq 0$ and $d_c\neq 0$ takes place
where the above free energy leads to the phase difference between  $d_b$ and  $d_{c}$ 
by $\pi/2$ when $\beta_2>0$.
It can be proved within the above free energy 
that third transition at $T=T^a_c$, which is designed to situate
a few mK below $T=T^c_c$
never realized at zero field.

Let us now discuss the phase diagrams for external field ${\bf 
H}$
applied parallel to and perpendicular to the $c$-axis.
We must take into account the vortex structure based on the 
above free enegy functional by adding the terms describing the 
spatially varied order parameter:
$F=F_{grad}+F_{bulk}$
with

$$F_{grad}=\sum_{j=a,b,c}\{K_1^j(|D_xd_j|^2+|D_yd_j|^2)
+K_2^j|D_zd_j|^2\}$$

$$F_{bulk}=\sum_{j=a,b,c}\alpha_0(T-T_{c}^j)|{d_j}|^2
+{1\over 2}\beta_1|{\bf d}|^4$$

$$+{1\over 2}\beta_2|{\bf d}^2|^2+{1\over 2}\Delta\chi_P |{\bf H}\cdot {\bf d}|^2.$$

\noindent
The gradient term $F_{grad}$  describes the spatial variation of the order parameter  under external field 
with the gauge invariant derivative $D_j=-i\hbar\partial_j-{2e \over c}A_j$ ($\bf A$  is the vector potential).
Here there exist two kinds ($K_1^j$ and $K_2^j$ ($j=a,b$ and $c$) of the gradient term associated with hexagonal symmetry where 
$K_1^b$ ($K_2^b$) can differ from $K_1^a$ and $K_1^c$ ($K_2^a$ and $K_2^c$) because of the symmetry breaking AF fluctuation,
that is, $K_1^b=K_1-\zeta {\bf M}^2, K_2^b=K_2-\zeta' {\bf M}^2$,
$K_1^a=K_1+\zeta {\bf M}^2,  K_2^a=K_2+\zeta' {\bf M}^2$,
$K_1^c=K'_1+\zeta {\bf M}^2,  K_2^c=K'_2+\zeta' {\bf M}^2$.
Here ${\bf M}^2$ is the amplitude 
of the AF fluctuation polarized along the $b$-axis.
It should be noted that there is no so-called 
gradient mixing term, which washes out 
the desired tetra-critical point,  in the present spin scenario.
This is quite 
different from that in the orbital scenarios\cite{sauls,joynt}
where the gradient mixing is inevitable.

The resulting phases in the $H$ vs $T$ shown in Fig.1 are 
fully characterized by specifying the $\bf d$ vector in each phase, that is,
its direction, the number of the components, and the relative 
phase.
The three vectors in the spin-triplet state are now
denoted by the real vectors $\hat{\bf a}$, $\hat{\bf b}$ and 
$\hat{\bf c}$ referring to the Cartesian coordinate in the hexagonal crystal. The high $T$ and low  (high)
field A(C) phase is described by a single component $\bf d$ vector, while the B phase in the low $T$ and low $H$ is a non-unitary state characterized by two-component $\bf d$ vector 
whose relative phase is ${\pi\over 2}$. 
Because of the $\Delta\chi_P$ term in the above free energy, which tends to align the ${\bf d}$-vector perpendicularly to ${\bf 
H}$, the B phase for ${\bf H}\parallel {\bf c}$ is further subdivided into $\hat{\bf b}$+i$\hat{\bf c}$ in low $H$ and $\hat{\bf b}$+i$\hat{\bf a}$ in high $H$. As $H(\parallel {\bf c})$ 
increases the $\bf d$ vector rotates,
corresponding to the KS behavior at $H\sim$2kG in Table 1.

We analyze the KS experiment by Tou {\it et al.}\cite{nmr,tou}, whose results are summarized in Table I:
In the case of ${\bf H}\parallel {\bf c}$ the KS does not (do) change below 
$T_c$ for $H\lesssim2$kG ($H>3$kG) (see their Fig.4\cite{nmr}).  This characteristic field $H_{rot}$ differs from that of the BC transition at 12kG, through which KS remains unchanged. According to the above theory within the 
B phase region $\hat{\bf b}$+i$\hat{\bf c}$ in low $H$ changes into $\hat{\bf b}$+i$\hat{\bf a}$ in high $H$ because the $\bf d$ 
vector tends to orient perpendicularly to the field direction, and 
the BC transition is from $\hat{\bf b}$+i$\hat{\bf a}$ to $\hat{\bf a}$.
Thus KS change does occur at $H_{rot}$, never at $H_{BC}$. 
This is precisely the case seen experimentally. The implication 
of the 
$\bf d$ vector rotation is obvious that SOC to lock the spin to 
the lattice is not strong, but weak enough to reoriente under a 
field as low as 2kG.

When ${\bf H}\parallel {\bf a}$, since both $\hat{\bf 
b}$+i$\hat{\bf c}$ (B phase) to $\hat{\bf c}$ (C phase) are 
perpendicular to $H$, KS should not change below $T_c$
for any field strengths, even crossing the BC boundary situated 
at $\sim$7kG. This coincides with the observations done for $H$=11, 4.7, 2.1 and 1.8KG (see their Fig.2\cite{nmr}).

As for ${\bf H}\parallel {\bf b}$, KS should and should not change below $T_c$
for the B phase and C phase respectively. The experiments under $H=$1.9, 2.4, 3.4, 4.7 and 5.3kG (B phase) and $H=$8.0 and 9.7kG (C phase) confirm this prediction (see their Fig.3\cite{nmr}). This means that 
the SOC in the basal plane is strong enough not to depin the $\bf d$ vector parallel to the the $b$-axis by the field up to $H_{BC}\sim$7kG.
In view of the difference in the characteristic energy scales for ${\bf H}\parallel {\bf b}$ ($T^b_c-T^c_c\sim$50mK) and ${\bf H}\parallel {\bf c}$ ($T^c_c-T^a_c\ll$50mK), this is understandable.

As pointed out by Tou {\it et al.}\cite{nmr}, their data allow to
draw the information on the A phase, which was quite scarce.
By carefuly comparing the two lowest field data sets for 
$H\parallel c$ (their Fig.4) and $H\parallel b$ (their Fig.3) one can clearly see that 
the decrease of KS does begin to start at $T_{c1}$ ($T_{c2}$)
for $H\parallel b$ ($H\parallel c$), indicating that
in the A phase the $\bf d$ vector points parallel to the 
$b$-axis. This is precisely consistent with our theory.

Importantly enough, the basal plane shows the two-fold 
symmetry.
The 45$^{\circ}$ direction data from the $a$-axis exhibits an intermediate decrease in the KS change. Theoretically we expect 
that the decrease  in $\chi$ corresponds to $\cos^2(\hat{\bf H}\cdot\hat{\bf d})={1\over 2}$.

The absolute values of the KS change ($\sim$0.06\% ($\parallel$${\bf b}$) vs $\sim$0.07\% ($\parallel$$ {\bf c}$)) which are proportional to the 
Pauli spin susceptibilities $\chi_i$ coming from the 
quasi-particles near the Fermi surface are roughly isotropic when 
taking into account the anisotropic hyperfine coupling constants 
($\sim20\%$ difference), supporting the spin space degeneracy scenario
(see their detailed analysis\cite{nmr}).
The magnitude of the estimated $\chi_i$ is extremely small; only 
a few \% of the total KS, but Tou's experimental accuracy can 
sensitively detect it, negating a persistent opinion that the KS 
measurement is meaningless because the susceptibility is 
dominated by the $T$-independent van Vleck term. 
It is indeed dominated by the van Vleck term, but they are 
undoubtedly probing the relevant quasi-particle contribution
or the Pauli susceptibility part.

An immediate consequence of this identification allows us to 
determine the direction of the spontaneously induced moment ${\bf M}_s$ associated with the non-unitarity in the B phase: ${\bf M}_s\propto i{\bf d}\times{\bf d}^{\ast}$$\parallel$$ {\bf a}$ at low 
fields where ${\bf d}=\hat{\bf b}$+i$\hat{\bf c}$. 
This spontaneous moment ${\bf M}_s$ was observed
and its magnitude was estimated  as $\sim$0.5G
by  $\mu$SR  experiment\cite{luke}. Now 
the direction of ${\bf M}_s$ should 
be probed. 
Since macroscopically this moment is cancelled out by the screening surface current and thus the magnetic induction $B=0$ in the bulk, the unscreened ${\bf M}_s$ survives only near the surface whose depth is an order of the penetration length.
If the supercurrent whose velocity is 
$v_s$ (say, 1cm/sec) flows along the $c$-axis, the electric field 
${\bf E}={\bf v}_s\times{\bf M}_s$ is spontaneously induced transversely, i.e., across the $b$-axis, whose order of magnitude is estimated as 
$\sim10^{-7}$volt/m. This  non-trivial prediction should be  checked 
experimentally.

Having analyzed Tou's experiment\cite{nmr,tou}, we discuss some of the 
critiques to our theory in the light of the new findings. The 
crossing phenomenon of the two upper critical field curves
for $H_{c2}\perp 
{\bf c}$ and $\parallel {\bf c}$ has been taken as strong evidence 
that  the Pauli limiting is absent (present) for the former 
(latter), supporting the E$_{2u}$ scenario\cite{choi}. Tou's experiment\cite{nmr,tou}
shows that this is not the case because both $\chi$'s 
remain unchanged below $T_c$.

Although we have determined the spin structure of the Cooper 
pair fairly completely, the precise gap topology remains to be 
seen. 
Recent several direction-resolved transport~\cite{taillefer} and  thermodynamic~\cite{tenya} measurements for each 
phase should be reexamined in the light of the present theory. We emphasize that the present scenario gives rise to different gap structures for the B, and A and C phases 
because the former (latter two) state is non-unitary (unitary), which have distinctively different low enegy excitation spectral structures.

In conclusion, we have identified the spin structures of the Cooper pair in each phase of UPt$_3$, which are fully consistent 
with the NMR experiment by Tou {\it et al.}\cite{nmr}. It is rather self-evident now that since the KS data show at least the four 
different behaviors for the principal three directions, it is impossible to understand in terms of the two-component pairing 
state such as the orbital scenarios.  
The main attractive force for the spin-triplet pairing comes from the paramagnetic fluctuations observed 
before\cite{bern}. The triple degeneracy is weakly broken by the
AF fluctuation, resulting in one of the complex $\bf d$ vector component locked to the $b$-axis. The fact that the remaining component of the $\bf d$ vector is rotatable 
under weak fields verifies experimentally that the effective 
spin-orbit coupling felt by Cooper pairs  is weak. 

The authors thank  H. Tou, Y. Kitaoka, Y. ${\bar{\rm O}}$nuki, T. 
Sakakibara and A. Sawada for useful experimental information. In 
particular, we are much indebted to H. Tou who has performed 
successfully this difficult NMR experiment and to his day-to-day 
information.

\vspace{.5cm}

\noindent
$\ast$Electronic address: machida@mp.okayama-u.ac.jp

\noindent
$\dagger$Electronic address: ohmi@ton.scphys.kyoto-u.ac.jp

\narrowtext
\begin{table}
\caption{Summary of the Knight shift experiment
for the principal directions $a$, $b$, and $c$-axes.
$\times$ ($\circ$) denotes the Knight shift unchanges (changes) below  $T_c$.}
\begin{tabular}{cccc}
&$a$&$b$&$c$\\
\tableline
high field&$\times$&$\times$&$\times$\\
\tableline
low field&$\times$&$\bigcirc$&$\bigcirc$\\
\end{tabular}
\label{table1}
\end{table}

\begin{figure}
\epsfysize=4.5cm
\epsfbox{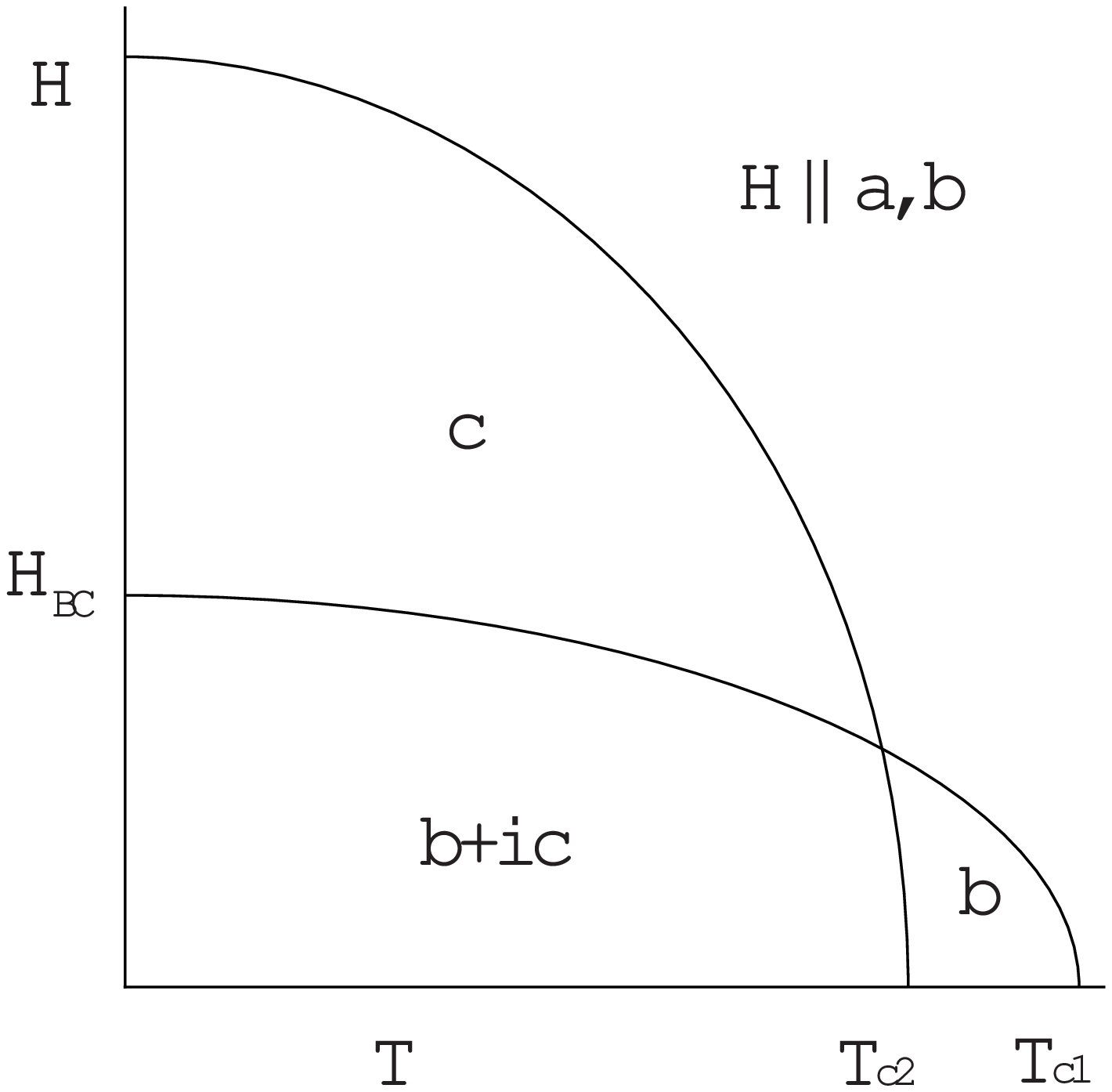}
\epsfysize=4.5cm
\epsfbox{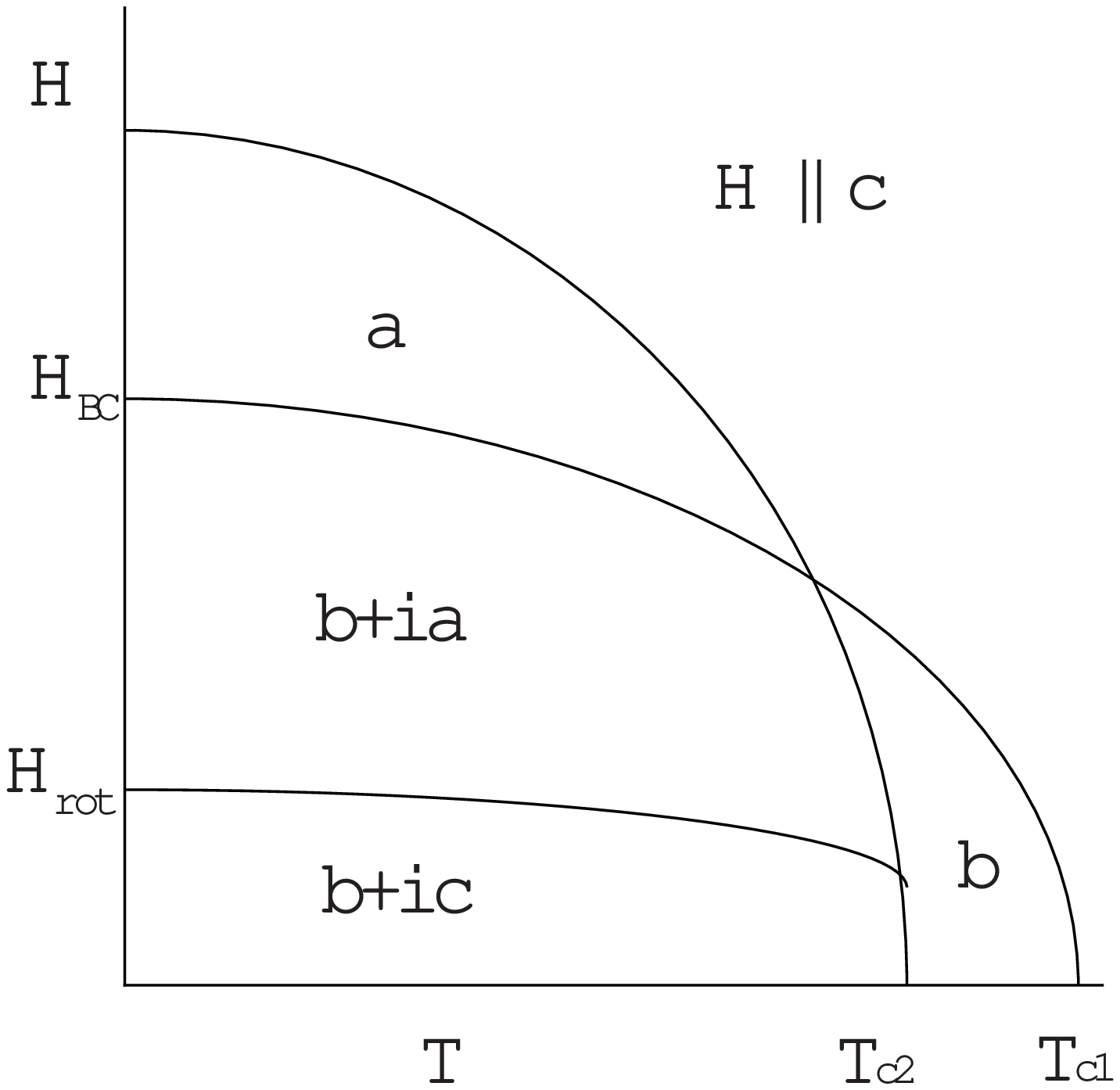}
\caption{
 Schematic phase diagrams in $H$ vs $T$ for $H\parallel a, b$ and $H\parallel c$ where the direction and the components of the $\bf d$ vector for each phase are indicated. ($H_{BC}$ the second phase transition between the B and C phases,
$H_{rot}$ the rotation field of the $\bf d$ vector)
}
\label{fig:1}
\end{figure}


\begin{thebibliography}{99}



\bibitem{heffner}R. H. Heffner and M. R. Norman, Comments Cond. 
Mat. Phys. {\bf 17}, 361 (1996). H. v. L\"ohneysen, Physica {\bf B 
197}, 551 (1994). M. Sigrist and K. Ueda, Rev. Mod. Phys. {\bf 63}, 239 (1991).






\bibitem{sauls}
J. A. Sauls, J. Low Tem. Phys. {\bf 95},153 (1994).
J. A. Sauls, Adv. Phys. {\bf 43}, 153 (1994).


\bibitem{joynt}K. A. Park and R. Joynt, Phys. Rev. Lett. {\bf 74}, 
4734 (1995).


\bibitem{our}K. Machida and M. Ozaki, Phys. Rev. Lett. {\bf 66}, 3293 (1991).
T. Ohmi and K. Machida, {\it ibid.} {\bf 71}, 625 (1993).
K. Machida, T. Ohmi and M. Ozaki, J. Phys. Soc. {\bf 62}, 3216 (1993). 
T. Ohmi and K. Machida, J. Phys. Soc. {\bf 65}, 4018 (1996). 

\bibitem{garg}A. Garg and D. Chen, Phys. Rev. B{\bf 49}, 479 (1994).




\bibitem{nmr}H. Tou, {\it et al.}, to be published in Phys. Rev. Lett..



\bibitem{tou}H. Tou, {\it et al.}, Phys. Rev. Lett. {\bf 77}, 1374 (1996).
Also see, Y. Kohori, {\it et al.}, J. Phys. Soc. Jpn. {\bf 56},  2263 (1987).



\bibitem{anderson}P. W. Anderson, Phys. Rev. B{\bf 30}, 4000 (1984).



\bibitem{aeppli3}G. Aeppli, {\it et al.},  Phys. Rev. Lett. {\bf 63},  676 (1989).

\bibitem{isaacs}E. D. Isaacs, {\it et al.}, Phys. Rev. Lett. {\bf 75}, 1178 (1995).



\bibitem{lussier}B. Lussier, {\it et al.}, Phys. Rev. B {\bf 54}, R6873 (1996). It is noted that to explain the observed isotropy within the basal plane the orbital scenarios assume the rotation of the AF moment so as to keep it perpendicularly to $\bf H$, which is not observed. They also notice that it is impossble to distinguish whether it is the triple $Q$ structure or the single $Q$ with three magnetic domains. As shown here the former must be the case to be consistent with the NMR experiment.



\bibitem{midgley}P. A. Midgley, {\it et al.}, Phys. Rev. Lett. {\bf 70}, 678 (1993). Also see, K. Elboussiri, Appl. Phys. A{\bf 59}, 223 (1994).
B. Ellman, {\it et al.}, Physica B{\bf 205}, 346 (1995). B. Ellman, {\it et al.}, preprint (cond-mat/9704125).




\bibitem{luke}G. M. Luke, {\it et al.}, Phys. Rev. Lett. {\bf 71}, 1446 (1993). Also see recent reports: D. A. Brawner, {\it et al.}, Physica B{\bf 230-232}, 338 (1997). P. Dalmas de R\'{e}otier, {\it et al.}, Phys. Lett. A{\bf 205}, 239 (1995).

\bibitem{choi}C. Choi and J. A. Sauls, Phys. Rev. Lett. {\bf 66}, 484 (1991).


\bibitem{taillefer}B. Ellman, {\it et al.}, Phys. Rev. B {\bf 54}, 9043 (1996).

\bibitem{tenya}K. Tenya, {\it et al.}, Phys. Rev. Lett. {\bf 77}, 3193 (1996).



\bibitem{bern}N. R. Bernhoeft and G. C. Lonzarich, J. Phys. C {\bf 7},  7325 (1995).





\end{thebibliography}
\end{document}